\documentstyle[prl,aps,preprint]{revtex} \tighten \begin{document} \draft
\title{Measurement of E2 transitions in the Coulomb dissociation of $^{8}$B}
\author{B.~Davids$^{1,2}$, D.W.~Anthony$^{1,3}$, Sam~M.~Austin$^{1,2}$,
D.~Bazin$^{1}$, B.~Blank$^{1,4}$, J.A.~Caggiano$^{1,2}$, M.~Chartier$^{1}$,
H.~Esbensen$^{5}$, P.~Hui$^{1}$\cite{mit}, C.F.~Powell$^{1,3}$,
H.~Scheit$^{1,2}$, B.M.~Sherrill$^{1,2}$, M.~Steiner$^{1}$, P.~Thirolf$^{\,6}$}
\address{$^{1}$ National Superconducting Cyclotron Laboratory, Michigan State
University, East Lansing, Michigan 48824\\$^{2}$ Department of Physics and
Astronomy, Michigan State University, East Lansing, Michigan 48824\\$^{3}$
Department of Chemistry, Michigan State University, East Lansing, Michigan
48824\\$^{4}$ Centre d'Etudes Nucl\'{e}aires de Bordeaux-Gradignan, F-33175
Gradignan Cedex, France\\$^{5}$ Physics Division, Argonne National Laboratory,
Argonne, Illinois 60439\\$^{6}$ Ludwig Maximilians Universit\"{a}t M\"{u}nchen,
Am Coulombwall 1, D-85748 Garching, Germany} \date{\today} \maketitle
\begin{abstract}In an effort to understand the implications of Coulomb
dissociation experiments for the determination of the
$^{7}$Be(\textit{p},$\gamma$)$^{8}$B reaction rate, longitudinal momentum
distributions of $^{7}$Be fragments produced in the Coulomb dissociation of 44
and 81 MeV/nucleon $^{8}$B beams on a Pb target were measured. These
distributions are characterized by asymmetries interpreted as the result of
interference between E1 and E2 transition amplitudes in the Coulomb breakup. At
the lower beam energy, both the asymmetries and the measured cross sections are
well reproduced by perturbation theory calculations, allowing a determination of
the E2 strength. This measurement yields S$_{E2}$/S$_{E1}$ = 6.7 $^{+2.8}_{-1.9}
\times 10^{-4}$ at the 0.63~MeV 1$^{+}$ resonance.\end{abstract}


The long-standing discrepancy between observed and predicted fluxes of neutrinos
from the sun has become known as the solar neutrino problem. All solutions to
this problem, including the possibility that solar neutrinos have mass and
change (oscillate) into different types of neutrinos to which current detectors
are less sensitive \cite{bahcall}, rely on accurate predictions of solar
neutrino fluxes. The rates of nuclear reactions that produce solar neutrinos
should be known well enough that they do not limit the precision of the
predicted fluxes. Of all the nuclear reaction rates that influence the flux of
high energy solar neutrinos, the rate of the radiative capture reaction
$^{7}$Be(\textit{p},$\gamma$)$^{8}$B is the least well known \cite{INT}. It is
customary to characterize its energy dependent cross section by a cross section
factor S$_{17}$(E), the value of which at E $\approx$ 0 determines the rate of
this reaction in the sun. Direct measurements of S$_{17}$(0) are difficult and
have yielded conflicting results
\cite{kavanagh60,parker,kavanagh69,vaughn,wiezorek,filippone,hammache}. A recent
analysis of data on S$_{17}$(0) \cite{INT} yielded a recommended value of
19$^{+4}_{-2}$~eV~b, a figure too imprecise for a detailed understanding of
solar neutrino experiments.

With the hope of providing a more reliable value, or at least a measurement with
different systematic errors, S$_{17}$(0) was measured indirectly at RIKEN by
studying the Coulomb dissociation of $^{8}$B \cite{motobayashi}. The cross
section of the reaction $^{208}$Pb($^{8}$B,$^{7}$Be$+$\textit{p})$^{208}$Pb was
obtained as a function of the relative energy of the $^{7}$Be and \textit{p}
fragments. In such an experiment, the heavy target nucleus creates a virtual
photon field in the rest frame of the incident $^{8}$B projectile. Virtual
photons with energies greater than or equal to 137.4 keV can dissociate the
$^{8}$B nucleus into $^{7}$Be + \textit{p}. The principle of detailed balance
then gives the radiative capture cross section for photons of a given
multipolarity. Thus a Coulomb dissociation measurement provides information
about the inverse radiative capture reaction rate. In the future, Coulomb
breakup experiments will be an important source of information about nuclear
reactions of astrophysical interest inaccessible by other means. Gaining an
understanding of the contributions of photons of different multipolarities and
their interferences will be crucial in the interpretation of these experiments.
To do so for the important $^{7}$Be(\textit{p},$\gamma$)$^{8}$B reaction is the
purpose of this letter.

A considerable controversy arose over the interpretation of the RIKEN data.
Although at solar energies the radiative capture reaction proceeds almost
exclusively by E1 induced transitions, E2 photons can contribute significantly
in the Coulomb dissociation process. The controversy centered on the role of E2
transitions in the RIKEN experiment. It was argued that the E2 contribution was
uncertain and could be large \cite{langanke,shyam96}; the E2 strength is
difficult to estimate reliably on theoretical grounds \cite{gai}. Two attempts
to measure the E2 strength have concluded that it is small
\cite{vonschwarzenberg,kikuchi}. However, as we will discuss later, it is not
certain that these measurements are free from the effects of background or
nuclear induced breakup.

Measurements of the distribution of longitudinal momenta of $^{7}$Be fragments
resulting from the breakup of $^{8}$B on heavy targets provide an independent
measurement of E2 strength. First order perturbation theory calculations of the
Coulomb dissociation of $^{8}$B predict that the distribution of the
longitudinal momenta of the emitted $^{7}$Be fragments will be asymmetric due to
interference between E1 and E2 transition amplitudes \cite{esbensen}. The
magnitude of this asymmetry depends on the beam energy, because the ratio of the
number of virtual E1 photons to E2 photons increases with beam energy. Hence, an
effective way to gauge the strength of E2 transitions in the Coulomb
dissociation of $^{8}$B is to measure the asymmetry of the longitudinal momentum
distributions of the emitted fragments at different beam energies. A recent
study of the longitudinal momentum distribution of $^{7}$Be fragments from the
Coulomb dissociation of 41~MeV/nucleon $^{8}$B on a gold target \cite{kelley}
found an asymmetry of roughly the predicted size, but poor statistics prevented
a definitive conclusion.

We used the new S800 spectrometer at the National Superconducting Cyclotron
Laboratory (NSCL) to carry out a much improved experiment. The large solid angle
(7$^{\circ} \times 10^{\circ}$), high resolution, and large momentum acceptance
(6\%) of the S800 made possible a high precision, high statistics measurement.
The beam energies were chosen to be 44 and 81 MeV/nucleon. The lower energy is
close to the energies of the earlier experiments of Refs.
\cite{motobayashi,kelley}.

In this experiment, $^{12}$C ions from the K1200 cyclotron at the NSCL bombarded
a 1.9~g~cm$^{-2}$ Be target and produced $^{8}$B nuclei by fragmentation
\cite{geissel}. The A1200 fragment separator \cite{A1200} purified the secondary
$^{8}$B beams. A thin plastic scintillator was placed near the exit of the A1200
for time of flight, rate, and transmission measurements. A 300 $\mu$m silicon
p-i-n diode detector at the target position of the S800 was used to periodically
monitor the transmission and composition of the secondary beams.

$^{8}$B nuclei were dissociated in a 28 mg~cm$^{-2}$ Pb target. The spectrometer
was set at 0$^{\circ}$ to detect $^{7}$Be fragments, and was operated in a
dispersion matched mode, so that the momentum spread of the incident beam did
not limit the momentum resolution of the spectrometer. The focal plane of the
S800 was instrumented with two position sensitive cathode readout drift chambers
(CRDCs) \cite{crdc}, a 16 segment ionization chamber, and 3 thick plastic
stopping scintillators. Energy loss signals were provided by the ionization
chamber, and the first scintillator was the source of total energy signals.
Reaction products were unambiguously identified by comparing the energies and
energy losses of the detected particles with those of a calibration beam of
$^{7}$Be having the same rigidity as the $^{8}$B beam. The ion optics code
\textsc{cosy infinity} \cite{cosy} was used to calculate momenta and scattering
angles for each event from the two dimensional position signals provided by each
CRDC and the magnetic field settings.

First order perturbation theory calculations of the Coulomb dissociation of
$^{8}$B on Pb at the energies of this experiment based on the model of Ref.
\cite{esbensen} have been performed. The model predicts  that S$_{17}$(0) = 17
eV~b, and that S$_{E2}$/S$_{E1}$ = 9.5$\times$10$^{-4}$ at the 0.63~MeV 1$^{+}$
resonance. This E2 strength is smaller than that of the model of Kim {\em et
al.} \cite{kim} by about a factor of 2. The total momentum distribution was
calculated for several $^{7}$Be scattering angle cuts. Projecting these events
on the beam direction yields the calculated longitudinal momentum distribution.

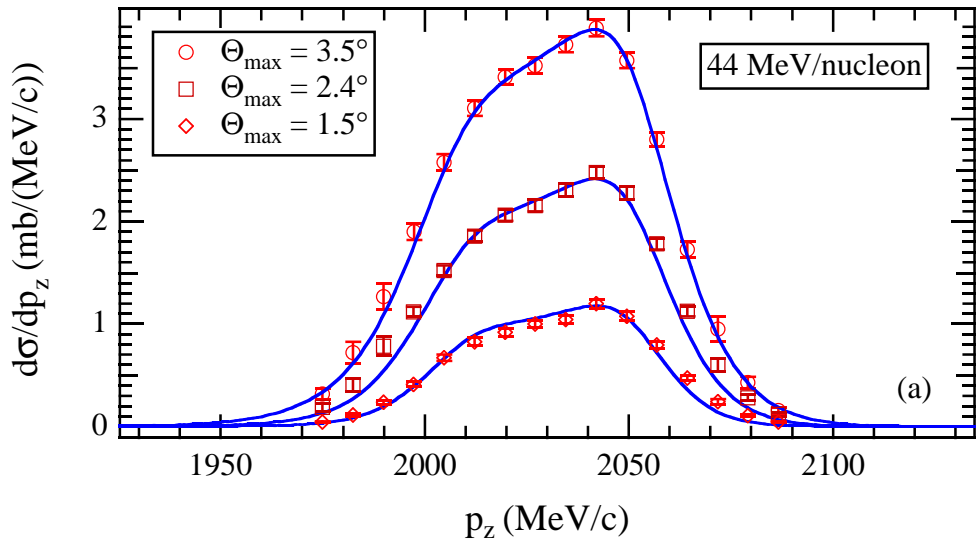
\begin{figure} \caption{(a) Laboratory frame longitudinal momentum distributions
of $^{7}$Be fragments formed in the Coulomb dissociation of 44~MeV/nucleon
$^{8}$B on Pb with maximum scattering angles of 1.5, 2.4, and 3.5$^{\circ}$. The
curves are the results of first order perturbation theory calculations
convoluted with the experimental resolution of 5~MeV/c. The error bars indicate
the relative uncertainties of the data points, which are dominated by
statistical errors. (b) Central region of the 3.5$^{\circ}$ angle cut at the
same beam energy. The curves are calculations performed with different E2
strengths, normalized to the center of the distribution.} \label{fig1}
\end{figure}

The measured longitudinal momentum distributions of $^{7}$Be fragments produced
in the Coulomb dissociation of $^{8}$B on Pb at 44 MeV/nucleon are shown in
Fig.\ \ref{fig1} (a) for three different angle cuts. The systematic
uncertainties in the measured cross sections are $\pm$ 10\% and include
contributions from the target thickness, acceptance and CRDC efficiency
corrections, and beam intensity, added in quadrature. The predicted longitudinal
momentum distributions, convoluted with the experimental resolution of 5~MeV/c,
are superposed on the measured distributions. The description of the data with
the original model was quite good, but not precise. It would be surprising if a
better prediction were obtained a priori. Both the value of S$_{17}$(0) and the
E2 strength are implicit in the structure model used, but are not robust
predictions of such models. Different models yield different predictions, as was
noted earlier. Since the predicted cross sections depend on S$_{17}$(0) and the
E2 strength, the normalization and E2 strength are effectively free parameters.
In addition, there is the 10\% systematic uncertainty in the normalization of
our cross section measurements. We therefore adjusted the E2 strength and the
overall normalization of the model to minimize $\chi^{2}$ for the central 6
points of the 3.5$^{\circ}$ distribution. This yields a normalization factor of
1.22 and an E2 strength 0.7 times as large as the original value. The value of
S$_{17}$(0) corresponding to this normalization factor is 21 eV~b, well within
the limits of the recommendation of Ref. \cite{INT}. The value of the ratio
S$_{E2}$/S$_{E1}$ at the 0.63~MeV 1$^{+}$ resonance corresponding to this E2
strength is 6.7 $\times 10^{-4}$, which is consistent with the upper limit of 7
$\times 10^{-4}$ given in Ref. \cite{gai}. It is worth noting that the E2
strength was extracted from the present experiment under the assumption of first
order perturbation theory. If higher order post-acceleration effects are large,
then the E2 strength extracted here is a lower limit. This is because for a
given E2 strength, the predicted asymmetry is smaller when higher order effects
are included than when they are neglected \cite{esbensen}.

Fig.\ \ref{fig1} (b) shows the central region of the 44 MeV/nucleon momentum
distribution for the 3.5$^{\circ}$ angle cut. Also shown are three calculations
with different E2 strengths, expressed as fractions of the original E2 strength
of the model, normalized to the center of the distribution. The dependence of
the calculated asymmetry on the E2 strength is apparent.

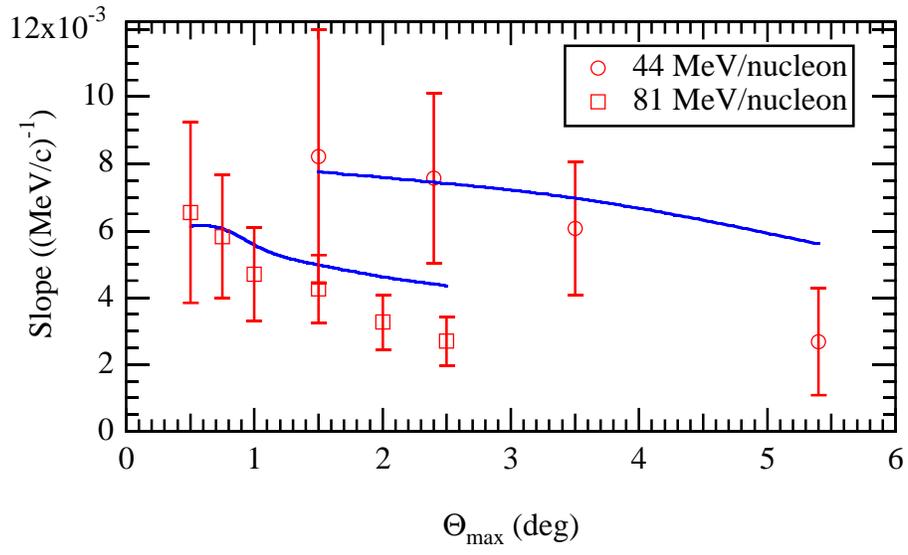
\begin{figure} \caption{Comparison of slopes extracted from the central regions
of the logarithms of the measured and theoretical longitudinal momentum
distributions plotted versus maximum scattering angle. The agreement between the
experimental results and the calculations is good at the smallest angle cuts.
Deviations between the experiment and the theory are evident at large scattering
angles, suggesting that nuclear breakup becomes important at these angles.}
\label{fig2} \end{figure}

In order to compare the asymmetries measured in the experiment with those
predicted by the model, the slopes of the central regions of the longitudinal
momentum distributions were extracted. However, the slopes of the theoretical
distributions are proportional to the normalization factor by which the
calculation has been multiplied. It is possible to eliminate this dependence on
the normalization factor by taking the logarithm of the distributions before
extracting a slope. Therefore, straight lines were fitted to the logarithms of
the measured and theoretical distributions between 2020 and 2035 MeV/c for the
lower energy, and between 2771 and 2791 MeV/c for the higher energy. The results
of this comparison are shown in Fig.\ \ref{fig2}. The calculations at both
energies were performed with the same optimal E2 strength described above. At
both energies, the experimental slopes decrease more rapidly with angle than the
Coulomb dissociation calculation predicts. This is interpreted as the result of
nuclear induced breakup at angles approaching the grazing angle, which is
4.4$^{\circ}$ at the higher beam energy. Nuclear breakup results in a symmetric
longitudinal momentum distribution \cite{kelley}. The calculations presented
here do not account for nuclear processes. The systematically smaller slopes at
the higher beam energy reflect the lesser relative importance of E2 transitions
there.

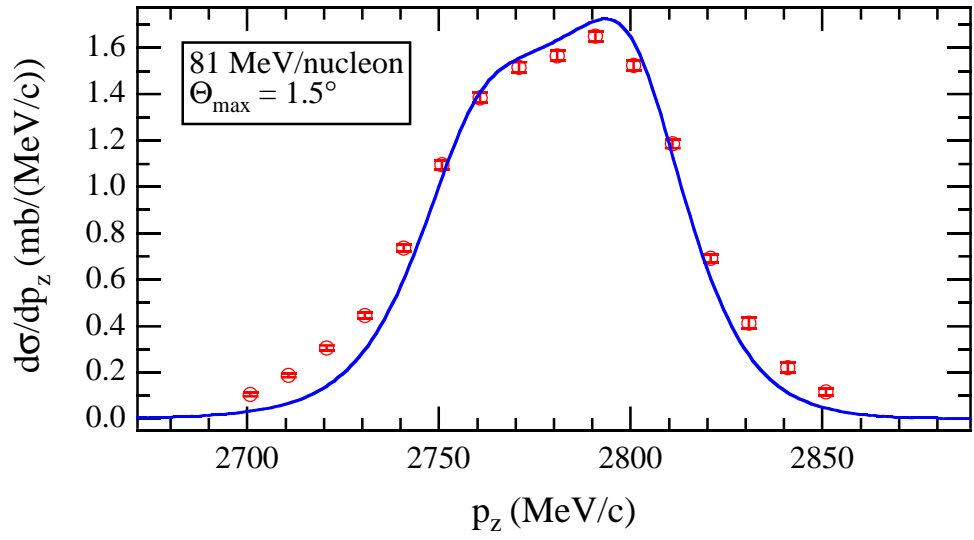
\begin{figure} \caption{Measured longitudinal momentum distribution of $^{7}$Be
fragments with scattering angles less than 1.5$^{\circ}$ formed in the
dissociation of 81 MeV/nucleon $^{8}$B, along with the prediction of the model.
The slightly greater width of the experimental distribution indicates the
presence of a broad component interpreted as the result of nuclear stripping not
accounted for in the calculation.} \label{fig3} \end{figure}

Fig.\ \ref{fig3} depicts a measured longitudinal momentum distribution of
$^{7}$Be fragments produced in the dissociation of $^{8}$B on Pb at 81
MeV/nucleon. Also shown is the prediction of the model with the same optimal E2
strength, again convoluted with the experimental resolution. The agreement
between experiment and calculation at this energy is not as good as at 44
MeV/nucleon. The main difference between the prediction of the model and the
experimental measurement is the slightly greater width of the measured
distribution. It is likely that nuclear stripping (not accounted for by the
model) is responsible for broadening the measured distributions. The nuclear
stripping cross sections are approximately equal at the two beam energies of
this experiment \cite{hencken}, while the Coulomb excitation cross section at a
given impact parameter is roughly proportional to the inverse square of the beam
velocity. Hence nuclear stripping is relatively more important at 81
MeV/nucleon, and it will probably not be possible to understand the measurement
at this energy without including nuclear processes in the calculation. For this
reason we utilized only the 44 MeV/nucleon data in extracting the E2 strength.

Good agreement between the observed and predicted shapes of the distributions
implies that the shapes of the E1 and E2 responses predicted by the model of
Ref. \cite{esbensen} are realistic, requiring only slight adjustments in
absolute magnitude. The evident asymmetry of the distributions, characteristic
of interference between $\ell$=1 and $\ell$=2 amplitudes, is therefore
interpreted as the result of interference between E1 and E2 transition
amplitudes. The asymmetry is observed at scattering angles corresponding to
minimum impact parameters greater than 25 fm, indicating that it arises from
Coulomb and not nuclear processes.

The momentum distributions measured in this experiment are consistent with the
low statistics measurement on the gold target reported in Ref. \cite{kelley}.
However, they apparently contradict the results of two experiments that found
the E2 component of the Coulomb dissociation of $^{8}$B to be smaller than the
theoretical predictions. The experiment of Ref. \cite{kikuchi} was designed to
determine the E2 contribution to the Coulomb dissociation of $^{8}$B by
measuring the angular distributions of the emitted fragments. While
acknowledging the possibility of large systematic errors \cite{kikuchi}, the E2
strength obtained was much smaller than all published theoretical predictions.
The interpretation of this experiment is complicated by the necessity of
extracting the breakup yield from a large background due to reactions in a
helium bag located between the target and detector systems. This background,
which varies with angle, is considerably larger than the yield from the target.
Although the experiment appears to have been done well, the subtraction of this
background, particularly from the large angle regime where E2 transitions are
important and the yield is small, could be problematic. Nuclear effects at large
angles (6-10$^{\circ}$) \cite{shyam96} also complicate the extraction of E2
strength.

In the experiment of Ref. \cite{vonschwarzenberg}, $^{7}$Be fragments from the
breakup of 26~MeV $^{8}$B on a nickel target were detected in detector
telescopes located at 45$^{\circ}$ with respect to the beam axis. E2 transitions
were expected to account for most of the breakup cross section. The upper limit
of the dissociation cross section reported in Ref. \cite{vonschwarzenberg} is
smaller than predictions based on any published model of Coulomb dissociation
\cite{shyam97}. Recent theoretical work suggests that nuclear processes may
contribute in the dissociation of $^{8}$B even at large impact parameters and
energies somewhat below the Coulomb barrier \cite{dasso}. In particular, the
authors of Ref. \cite{dasso} find that at the energy of the experiment of Ref.
\cite{vonschwarzenberg}, Coulomb and nuclear amplitudes interfere destructively
around 40$^{\circ}$. The size of the nuclear amplitude at this angle is
comparable to that of the Coulomb amplitude, so that the resulting total cross
section is about a factor of 4 smaller than the Coulomb dissociation cross
section. Since the measurement of Ref. \cite{vonschwarzenberg} was made at and
near this angle, the measured cross section may not be a good measure of the E2
strength.

In summary, longitudinal momentum distributions of $^{7}$Be fragments formed in
the Coulomb dissociation of 44 and 81~MeV/nucleon $^{8}$B on Pb have been
measured with high precision and statistics. At 44 MeV/nucleon, the shapes of
the measured distributions and the magnitudes of the measured cross sections
agree with the predictions of first order perturbation theory calculations based
on a simple potential model description of the structure of $^{8}$B. The model
predicts that interference between E1 and E2 transition amplitudes will produce
observable asymmetries in the longitudinal momentum distributions of the
fragments formed in the Coulomb dissociation of $^{8}$B at the beam energies
investigated in this experiment. These asymmetries were in fact observed. The 81
MeV/nucleon distributions were not as well fit, perhaps because of a significant
nuclear component, so the 44 MeV/nucleon distributions alone were used to
extract the E2 strength. However, the theoretical slopes based on this E2
strength agree well with the 81 MeV/nucleon data at small angles, where nuclear
effects are small. The measured distributions are consistent with a value of 6.7
$^{+2.8}_{-1.9} \times 10^{-4}$ for the ratio S$_{E2}$/S$_{E1}$ at the 0.63~MeV
1$^{+}$ resonance. The calibration of the magnitude of the E1 response of the
model made in order to reproduce the measured cross sections implies a value for
S$_{17}$(0) that agrees with the recommendation of Ref. \cite{INT}. However, the
longitudinal momentum distribution of $^{7}$Be fragments is not the most
sensitive probe of this quantity. We hope that a careful measurement of the
$^{8}$B decay energy spectrum at low excitation energies, planned later this
year, will allow a more precise determination of S$_{17}$(0).

This work was supported by the U.S. National Science Foundation. One of us
(H.E.) was supported by the U.S. Department of Energy, Nuclear Physics Division,
under contract No. W-31-109-ENG-38.

\begin{references}

\bibitem[*]{mit}Present address: Lincoln Laboratory, Massachusetts Institute of
Technology, 244 Wood Street, Lexington, MA 02173. \bibitem{bahcall}J.N.~Bahcall,
{\em Neutrino Astrophysics} (Cambridge University Press, Cambridge, 1989).
\bibitem{INT}INT Workshop on Solar Nuclear Fusion Cross Sections, 1997, Rev.
Mod. Phys. (to be published). \bibitem{kavanagh60}R.W.~Kavanagh, Nucl. Phys.
{\bf 15}, 411 (1960). \bibitem{parker}P.D.~Parker, Phys. Rev. {\bf 150}, 851
(1966).\bibitem{kavanagh69}R.W.~Kavanagh {\em et al.}, Bull. Am. Phys. Soc. {\bf
14}, 1209 (1969). \bibitem{vaughn}F.J.~Vaughn {\em et al.}, Phys. Rev. C {\bf
2}, 1657 (1970). \bibitem{wiezorek}C.~Wiezorek {\em et al.}, Z. Phys. A {\bf
282}, 121 (1977). \bibitem{filippone}B.W.~Filippone, A.J.~Elwyn, C.N.~Davids,
and D.D.~Koetke, Phys. Rev. Lett. {\bf 50}, 412 (1983); Phys. Rev. C {\bf 28},
2222 (1983). \bibitem{hammache}F.~Hammache {\em et al.}, Phys. Rev. Lett. {\bf
80}, 928 (1998). \bibitem{motobayashi}T. Motobayashi {\em et al.}, Phys. Rev.
Lett. {\bf 73}, 2680 (1994). \bibitem{langanke}K.~Langanke and T.D.~Shoppa,
Phys. Rev. C {\bf 49}, R1771 (1994); Phys. Rev. C {\bf 51}, 2844 (1995); Phys.
Rev. C {\bf 52}, 1709 (1995). \bibitem{shyam96}R.~Shyam, I.J.~Thompson, and
A.K.~Dutt-Mazumder, Phys. Lett. B {\bf 371}, 1 (1996). \bibitem{gai}M.~Gai and
C.A.~Bertulani, Phys. Rev. C {\bf 52}, 1706 (1995).
\bibitem{vonschwarzenberg}J.~von~Schwarzenberg {\em et al.}, Phys. Rev. C {\bf
53}, R2598 (1996). \bibitem{kikuchi}T.~Kikuchi {\em et al.}, Phys. Lett. B {\bf
391}, 261 (1997). \bibitem{esbensen}H.~Esbensen and G.F.~Bertsch, Phys. Lett. B
{\bf 359}, 13 (1995); Nucl. Phys. A {\bf 600}, 37 (1996).
\bibitem{kelley}J.H.~Kelley {\em et al.}, Phys. Rev. Lett. {\bf 77} 5020 (1996).
\bibitem{geissel}H.~Geissel, G.~M\"{u}nzenberg, and K.~Riisager, Annu. Rev.
Nucl. Part. Sci. {\bf 45}, 163 (1995). \bibitem{A1200}B.M.~Sherrill {\em et
al.}, Nucl. Instrum. Methods B {\bf 70}, 298 (1992). \bibitem{crdc}J.~Yurkon
{\em et al.}, {\em National Superconducting Cyclotron Laboratory Annual Report},
207 (1996). \bibitem{cosy}M.~Berz {\em et al.}, Phys. Rev. C {\bf 47}, 537
(1993). \bibitem{kim}K.H.~Kim, M.H.~Park, and B.T.~Kim, Phys. Rev. C {\bf 35},
363 (1987). \bibitem{hencken}K.~Hencken, G.~Bertsch, and H.~Esbensen, Phys. Rev.
C {\bf 54}, 3043 (1996). \bibitem{shyam97}R.~Shyam and I.J.~Thompson, Phys.
Lett. B {\bf 415}, 315 (1997). \bibitem{dasso}C.H.~Dasso, S.M.~Lenzi, and
A.~Vitturi, to be published.


\end{references}
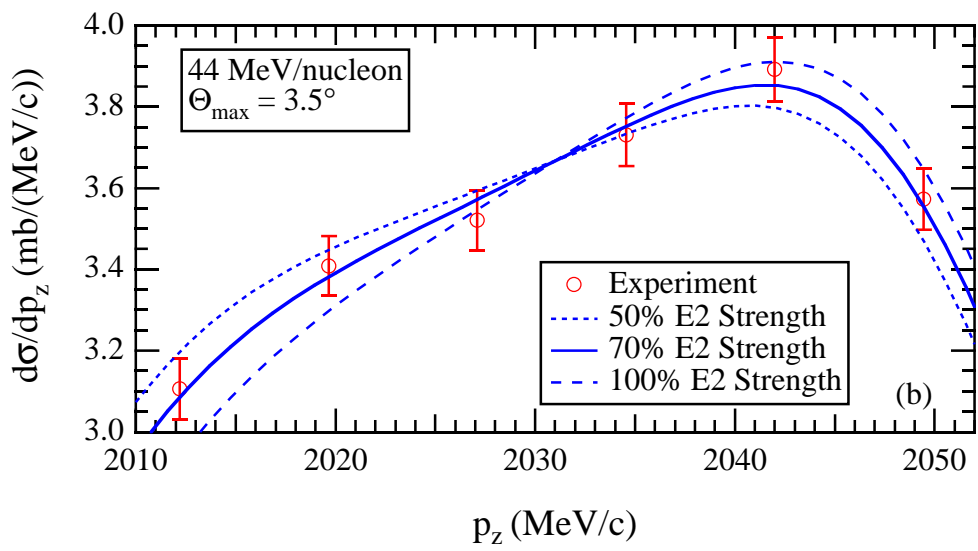
\end{document}